\documentclass[preprint,12pt]{elsarticle}

\usepackage{amssymb}

\usepackage{amsmath,amsthm,amssymb,mathrsfs}
\usepackage{bm}

\journal{Journal of Magnetism and Magnetic Materials}

\begin{document}

\begin{frontmatter}




\title{Role of additional microwave voltage on phase locking in voltage-controlled parametric oscillator}


\author{Tomohiro Taniguchi 
}


\address{
 National Institute of Advanced Industrial Science and Technology (AIST), Research Center for Emerging Computing Technologies, Tsukuba, Ibaraki 305-8568, Japan
}



\begin{abstract}
A demonstration of parametric oscillation of magnetization in nanostructured ferromagnets via voltage-controlled magnetic anisotropy (VCMA) effect provided an alternative approach to spintronic oscillator applications with low-energy consumption. 
However, the phase of this voltage-controlled parametric oscillator was unable to be locked uniquely by microwave VCMA effect. 
The oscillation phase is locked in one of two possible states, which originates from the fact that the frequency of the microwave voltage is twice that of the magnetization oscillation. 
In this work, we investigate the phase locking by additional microwave voltage through analytical and numerical studies of the Landau-Lifshitz-Gilbert equation. 
An analytical study suggests that the additional voltage makes one of two phases more stable than the other by having asymmetric potential for the phase. 
The simulation results indicate a trigonometric-function-like dependence of the locked rate on the phase of the additional voltage, which qualitatively agrees with the analytical theory and also suggests a possibility to manipulate the phase by the additional voltage. 
\end{abstract}


\end{frontmatter}





\section{Introduction}
\label{sec:Introduction}

Parametric oscillation is a phenomenon, where a periodic modulation of a system parameter causes an oscillation as in the case of a swing \cite{landau82,arnold89}. 
The parametric oscillation has been of interest not only from fundamental aspect but also from the viewpoint of practical applications such as logic devices based on parametrons made of electrical circuit \cite{goto59}. 
Recently, the parametric oscillation was observed in nanostructured ferromagnets \cite{yamamoto20} caused by microwave voltage-controlled magnetic anisotropy (VCMA) effect \cite{weisheit07}. 
The VCMA effect modulates a parameter of the ferromagnets, namely the perpendicular magnetic anisotropy field, by the modulation of electron states \cite{duan08,nakamura09,tsujikawa09} and/or induction of magnetic moment \cite{miwa17} near the ferromagnetic metal/insulating layer interface. 
The parametric oscillation was excited when the frequency $f$ of the microwave voltage is twice the Larmor frequency $f_{\rm L}=\gamma H_{\rm appl}/(2\pi)$ of the magnetization oscillation around an in-plane applied magnetic field $H_{\rm appl}$ ($\gamma$ is the gyromagnetic ratio). 
In this work, we call this oscillator the voltage-controlled parametric oscillator, for simplicity. 
This parametric oscillation was originally proposed to achieve a stable magnetization switching in non-volatile memory by the VCMA effect even in a long voltage-pulse regime \cite{yamamoto20}. 
The stability of the oscillation, however, may lead to a wide range of applicability for this parametric oscillation to be used in practical devices. 
In particular, the fact that the oscillation is driven by the VCMA effect may possibly be advantageous to other ferromagnetic oscillators, such as spin-torque oscillators driven by electric current \cite{kiselev03,rippard04,houssameddine07,kubota13}, because VCMA can significantly reduce the Joule heating caused by the electric current. 


The parametric oscillators often have a characteristic of not being able to lock the oscillation phase uniquely \cite{goto59}.  
This is in contrast to a unique phase locking in nonlinear oscillators by injecting periodic input \cite{pikovsky03}. 
The non-uniqueness of the phase originates from the fact that the frequency $f$ of the parameter is twice the value of the oscillator frequency. 
As a result, there are two possible values of the locked phase, and realizing which state depends on the initial state. 
This is also true for the voltage-controlled parametric oscillator \cite{taniguchi22,taniguchi23}. 
A similar phenomenon was recently observed even in an injection-locked spin-torque oscillator \cite{phan24}.  
The non-uniqueness of the oscillator phase is not suitable for certain practical applications, such as oscillator-based brain-inspired computing \cite{torrejon17,kudo17,riou19,prasad22,rodrigues23,imai23} but may probably be of interest for other 
Therefore, it is of great importance to investigate a method to manipulate the phase of the voltage-controlled parametric oscillator. 
In the case of the parametron made of electrical circuit \cite{goto59}, for example, an injection of output signal from one parametron to others causes a unique phase locking among them. 
It implies that an injection of an additional signal may be useful for manipulating the phase locking of parametric oscillators. 


In this work, we study a possibility of the phase locking in the voltage-controlled parametric oscillator by applying an additional microwave voltage. 
The frequency $f_{\rm a}$ of the additional microwave voltage differs from that ($f=2f_{\rm L}$) of the original microwave voltage driving the parametric oscillation. 
We solve the Landau-Lifshitz-Gilbert (LLG) equation numerically for various initial conditions and find that the VCMA effect due to the additional voltage changes the locked rate of the oscillator phase. 
The locked rate shows a trigonometric-function-like dependence with respect to the phase of the additional voltage. 
An analytical study on the LLG equation implies that the existence of two possible phases of the oscillator can be explained in terms of a potential energy for the phase having two equivalent stable states and the additional voltage results in asymmetric potential, which qualitatively agrees with the numerical results. 
These results provide a new method to manipulate the voltage-controlled parametric oscillators. 



\begin{figure}
\centerline{\includegraphics[width=0.5\columnwidth]{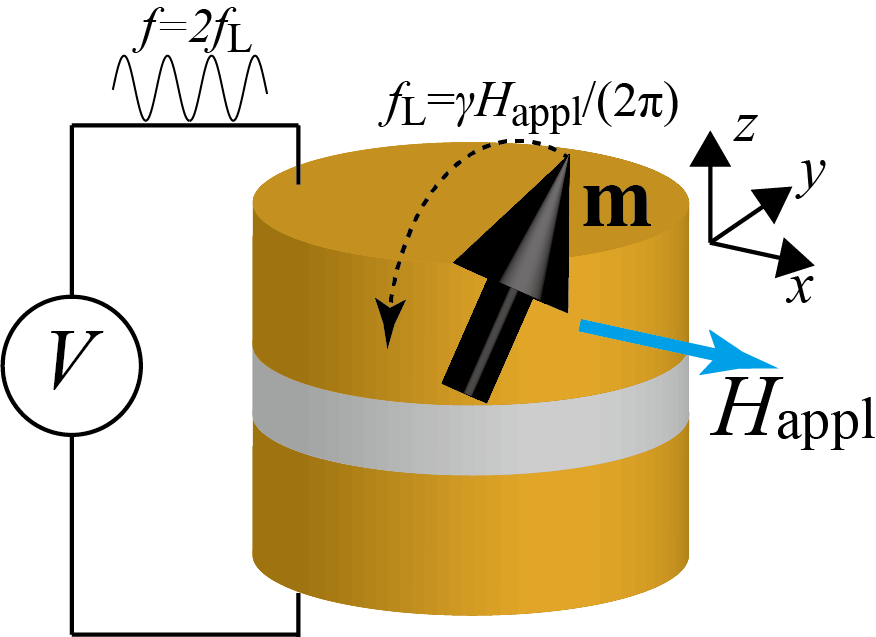}}
\caption{
            Schematic illustration of a voltage-controlled parametric oscillator. 
            A unit vector pointing in the magnetization direction in the free layer is denoted as $\mathbf{m}$. 
            In the parametric oscillation state, an oscillating voltage with a frequency $f=2f_{\rm L}$ drives a sustainable oscillation of the magnetization around an in-plane applied magnetic field $H_{\rm appl}$ with the Larmor frequency $f_{\rm L}=\gamma H_{\rm appl}/(2\pi)$. 
         \vspace{-3ex}}
\label{fig:fig1}
\end{figure}


\section{System description}
\label{sec:System description}

In Fig. \ref{fig:fig1}, we show a schematic illustration of a ferromagnetic/nonmagnetic/ferromagnetic trilayer, where the top and bottom ferromagnets correspond to free and reference layers, respectively. 
The $z$ axis points normal to the film plane, while the $x$ axis is parallel to the direction of the in-plane applied magnetic field. 
The unit vector pointing in the direction of the magnetization is denoted as $\mathbf{m}$, where we use the macrospin assumption whose validity has been confirmed for typical VCMA experiments \cite{yamamoto20}. 
The magnetization dynamics in the free layer is described by the LLG equation \cite{landau80,gilbert04}, 
\begin{equation}
  \frac{d \mathbf{m}}{dt}
  =
  -\gamma 
  \mathbf{m}
  \times
  \mathbf{H}
  +
  \alpha 
  \mathbf{m}
  \times
  \frac{d \mathbf{m}}{dt}, 
  \label{eq:LLG}
\end{equation}
where $\alpha$ is the Gilbert damping constant. 
The magnetic field $\mathbf{H}$ is 
\begin{equation}
  \mathbf{H}
  =
  H_{\rm appl}
  \mathbf{e}_{x}
  +
  H_{\rm K}
  m_{z}
  \mathbf{e}_{z}, 
  \label{eq:field_def}
\end{equation}
where $\mathbf{e}_{k}$ ($k=x,y,z$) is the unit vector pointing in the $k$ direction. 
The perpendicular magnetic anisotropy field $H_{\rm K}$ consists of the demagnetization (shape magnetic anisotropy) field, interfacial magnetic anisotropy field \cite{shinjo79,yakata09,ikeda10,kubota12}, and the contribution from the VCMA effect \cite{duan08,nakamura09,tsujikawa09,miwa17,maruyama09,shiota09,nozaki12,nozaki19}. 
Summarizing them, $H_{\rm K}$ can either be positive or negative by the VCMA effect; see also \ref{sec:AppendixA}. 
The value of $H_{\rm K}$ in this work is controlled as follows. 



\begin{figure}
\centerline{\includegraphics[width=1.0\columnwidth]{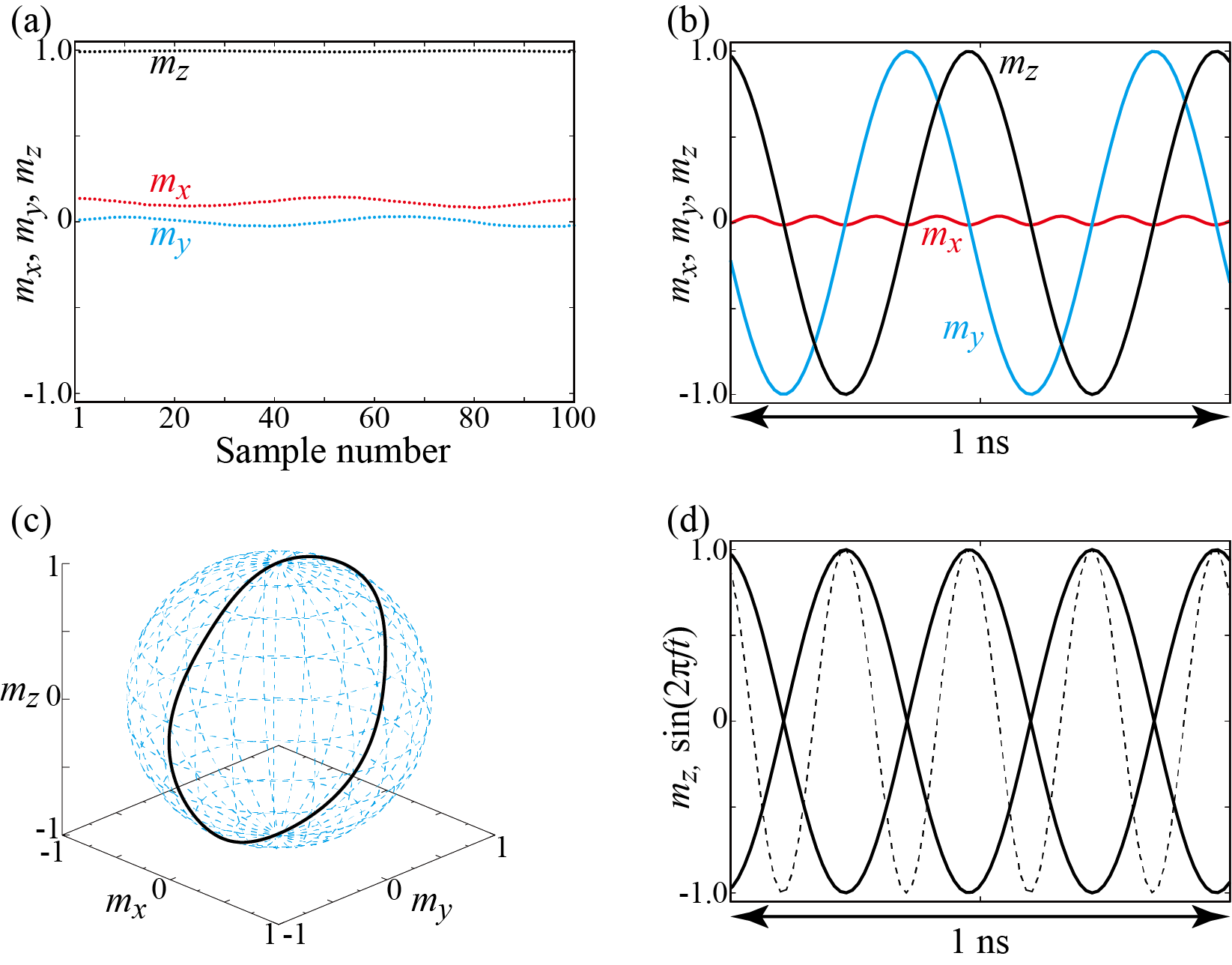}}
\caption{
            (a) $100$ samples of an initial state of $\mathbf{m}$ prepared by solving the LLG equation with thermal fluctuation. 
            (b) An example of time evolution of $m_{x}$ (red), $m_{y}$ (blue), and $m_{z}$ (black) in the parametric oscillation state. 
                An additional voltage generating $H_{\rm K}^{(2)}$ is set to be zero, for simplicity. 
            (c) Dynamical trajectory of the magnetization in the parametric oscillation state. 
            (d) Two examples of $m_{z}$ with different initial conditions (solid lines). 
                 A trigonometric function, $\sin(2\pi ft)$ with $f=2f_{\rm L}$, representing the oscillation of the alternating voltage, is also shown by a dotted line. 
         \vspace{-3ex}}
\label{fig:fig2}
\end{figure}



Firstly, we prepare $100$ natural initial conditions by the following method \cite{imai22}. 
We assume that, before applying voltage, the magnetization stays near an energetically stable state, $\mathbf{m}^{(0)}=[(H_{\rm appl}/H_{\rm Kd}),0,+\sqrt{1-(H_{\rm appl}/H_{\rm Kd})^{2}}]$, with a small-amplitude oscillation due to the thermal fluctuation, where $H_{\rm Kd}$ is the value of $H_{\rm K}$ in the absence of the voltage and is set to be $6.283$ kOe, according to Ref. \cite{yamamoto20}. 
The thermal fluctuation provides a random torque, $-\gamma\mathbf{m}\times\mathbf{h}$, whose components obey the fluctuation-dissipation theorem \cite{brown63}, 
\begin{equation}
  \langle h_{k}(t) h_{\ell}(t^{\prime}) \rangle 
  =
  \frac{2\alpha k_{\rm B}T}{\gamma MV}
  \delta_{k\ell}
  \delta(t-t^{\prime}), 
  \label{eq:FDT}
\end{equation}
where the temperature $T$, saturation magnetization $M$, and volume $V$ are set to be $T=300$ K, $M=995$ emu/cm${}^{3}$, and $V=\pi r^{2}d$, where the radius $r$ and the thickness $d$ are $50$ nm and $1.1$ nm, respectively \cite{yamamoto20}. 
The values of $H_{\rm appl}$, $\gamma$, and $\alpha$ are $720$ Oe, $1.764\times 10^{7}$ rad/(Oe s), and $0.005$. 
Solving the LLG equation with these parameters and the random torque, we pick up $100$ temporal solutions, which are summarized in Fig. \ref{fig:fig2}.(a); see also \ref{sec:AppendixB}.


Next, we investigate the solution of Eq. (\ref{eq:LLG}) in the presence of the VCMA effect. 
In the typical voltage-controlled parametric oscillator, both the direct and alternating voltages are applied \cite{yamamoto20}. 
The VCMA effect caused by the direct voltage makes the direct component of $H_{\rm K}$ close to zero because, if it remains finite, an instantaneous frequency of the magnetization oscillation becomes nonuniform and causes non-preferable error of the switching.  
Accordingly, $H_{\rm K}$ in this study has the alternating components only, where $H_{\rm K}$ is given by 
\begin{equation}
  H_{\rm K}
  =
  H_{\rm K}^{(1)}
  \sin(2\pi ft)
  +
  H_{\rm K}^{(2)}
  \sin(2\pi f_{\rm a}t + \delta), 
  \label{eq:H_K}
\end{equation}
where $H_{\rm K}^{(1)}$ is the amplitude of the oscillating perpendicular magnetic anisotropy field generated by the microwave VCMA effect with $f=2f_{\rm L}$. 
This perpendicular magnetic anisotropy field is the origin of the parametric oscillation \cite{yamamoto20}. 
In addition to this, we introduce another oscillating perpendicular magnetic anisotropy field with the amplitude $H_{\rm K}^{(2)}$, frequency $f_{\rm a}$, and phase $\delta$. 
We solve Eq. (\ref{eq:LLG}) with Eq. (\ref{eq:H_K}) for $100$ initial conditions. 


Before ending this section, let us confirm the excitation of the parametric oscillation by the VCMA effect in the absence of $H_{\rm K}^{(2)}$ \cite{yamamoto20} and show the presence of two possible locked phases. 
Figure \ref{fig:fig2}(b) shows an example of the solution $\mathbf{m}$ [$m_{x}$ (red), $m_{y}$ (blue), and $m_{z}$ (black)] of the LLG equation in a steady state, while Fig. \ref{fig:fig2}(c) shows the dynamical trajectory of the oscillation. 
The value of $H_{\rm K}^{(1)}$ is $300$ Oe. 
We see that the magnetization oscillates around the $x$ axis with approximately $m_{x}\simeq 0$. 
The oscillation frequency is confirmed as $2.02$ GHz, which equals to the Larmor frequency $f_{\rm L}$. 
In addition, we confirm the presence of two possible phases in the solution of the LLG equation; see Fig. \ref{fig:fig2}(d), where $m_{z}$ with two different initial conditions are shown by solid lines, in addition to the oscillation $\sin(2\pi ft)$ representing the oscillation of the microwave voltage (dotted lines). 
The solutions of the LLG equation finally saturate to one of these two states, depending on the initial state. 
In the following, we study how the additional VCMA effect ($\propto H_{\rm K}^{(2)}$) affects this oscillation phase. 



\section{Numerical simulation}
\label{sec:Numerical simulation}

In this section, we show the results of the numerical simulation with regard to the relation between the additional microwave VCMA effect and the phase locking. 
As shown in Fig. \ref{fig:fig2}(d), there are two possible phases of the magnetization oscillation. 
Therefore, we denote the number of the samples having one kind of the phases as $N_{+}$, while the number of samples having the other phase as $N_{-}$. 
In the parametric oscillation states, $N_{+}+N_{-}$ equals to the sample numbers, which is $100$ in this study. 
Then, we define a locked rate ${\rm LR}$ as 
\begin{equation}
  {\rm LR}
  =
  \frac{{\rm max}[N_{+},N_{-}]}{N_{+}+N_{-}}. 
  \label{eq:locked_rate}
\end{equation}
The locked rate is $1$ when the phase of the magnetization oscillation becomes independent of the initial state and is uniquely locked, while it is $0.5$ when two possible phases are equally realized. 
Before studying the role of the additional VCMA effect, we mention that the locked rate unnecessarily becomes $0.5$ for the conventional parametric oscillator. 
Although the phase depends on the initial state, the locked rate might be biased to one of two possible values because the possible initial states are concentrated near the energetically stable state, as shown in Fig. \ref{fig:fig2}(a), and thus, the difference in the initial states might be small. 
At the same time, however, we should note that even a slight difference of the initial state sometimes leads to a different phase, as shown in Fig. \ref{fig:fig2}(d). 
This phase difference is uncontrollable because the initial state is randomly distributed. 
The aim of the following calculation is to manipulate the phase by overcoming this randomness.


\begin{figure}
\centerline{\includegraphics[width=1.0\columnwidth]{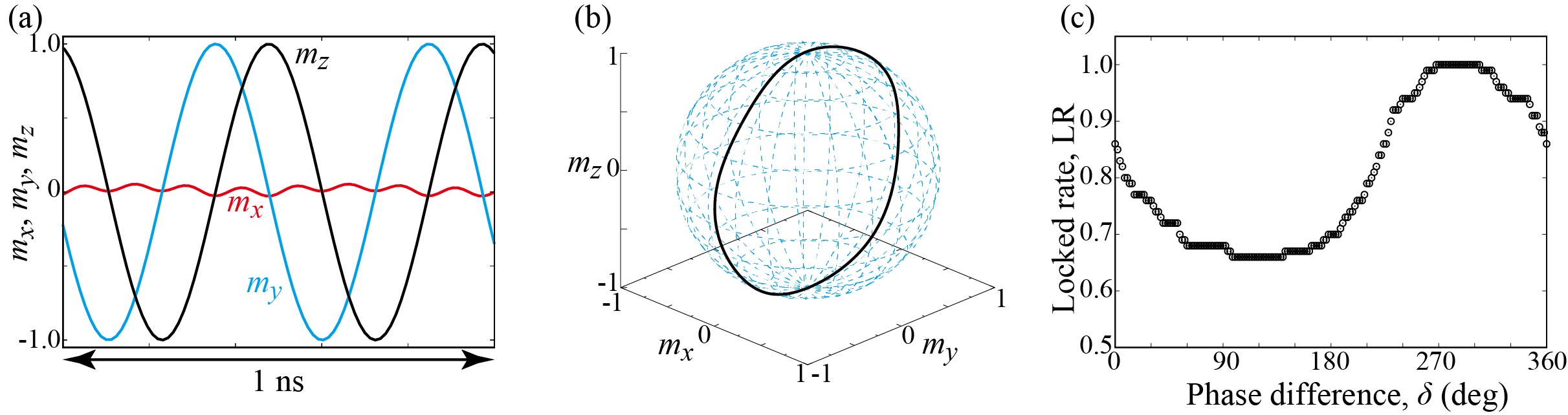}}
\caption{
            (a)  An example of time evolution of $m_{x}$ (red), $m_{y}$ (blue), and $m_{z}$ (black) in the parametric oscillation state. 
                The values of the parameters of an additional perpendicular magnetic anisotropy field are $H_{\rm K}^{(2)}=50$ Oe, $f_{\rm a}=f_{\rm L}$, and $\delta=0^{\circ}$.  
            (b) Dynamical trajectory of the magnetization in the presence of the additional perpendicular magnetic anisotropy. 
            (c) Locked rate as a function of $\delta$ for $H_{\rm K}^{(2)}=50$ Oe and $f_{\rm a}=f_{\rm L}$. 
         \vspace{-3ex}}
\label{fig:fig3}
\end{figure}


We first confirm that the additional VCMA effect does not significantly change the parametric oscillation. 
Figures \ref{fig:fig3}(a) and \ref{fig:fig3}(b) show an example of the magnetization dynamics in the presence of the additional microwave VCMA effect and the dynamical trajectory, where $H_{\rm K}^{(2)}=50$ Oe, $f_{\rm a}=f_{\rm L}$, and $\delta=0^{\circ}$. 
Comparing them with Figs. \ref{fig:fig2}(b) and \ref{fig:fig2}(c), we find that the parametric oscillation with the Larmor frequency is still excited. 
We should emphasize here, however, that the additional microwave VCMA effect plays a role in the phase locking. 
To clarify this point, we evaluate the locked rate for various $\delta$. 
The result is summarized in Fig. \ref{fig:fig3}(c), which implies that the locked rate changes with $\delta$, and its dependence resembles the behavior of a trigonometric function. 
Therefore, we conclude that the additional VCMA effect can be used to manipulate the phase of the voltage-controlled parametric oscillator. 



\begin{figure}
\centerline{\includegraphics[width=1.0\columnwidth]{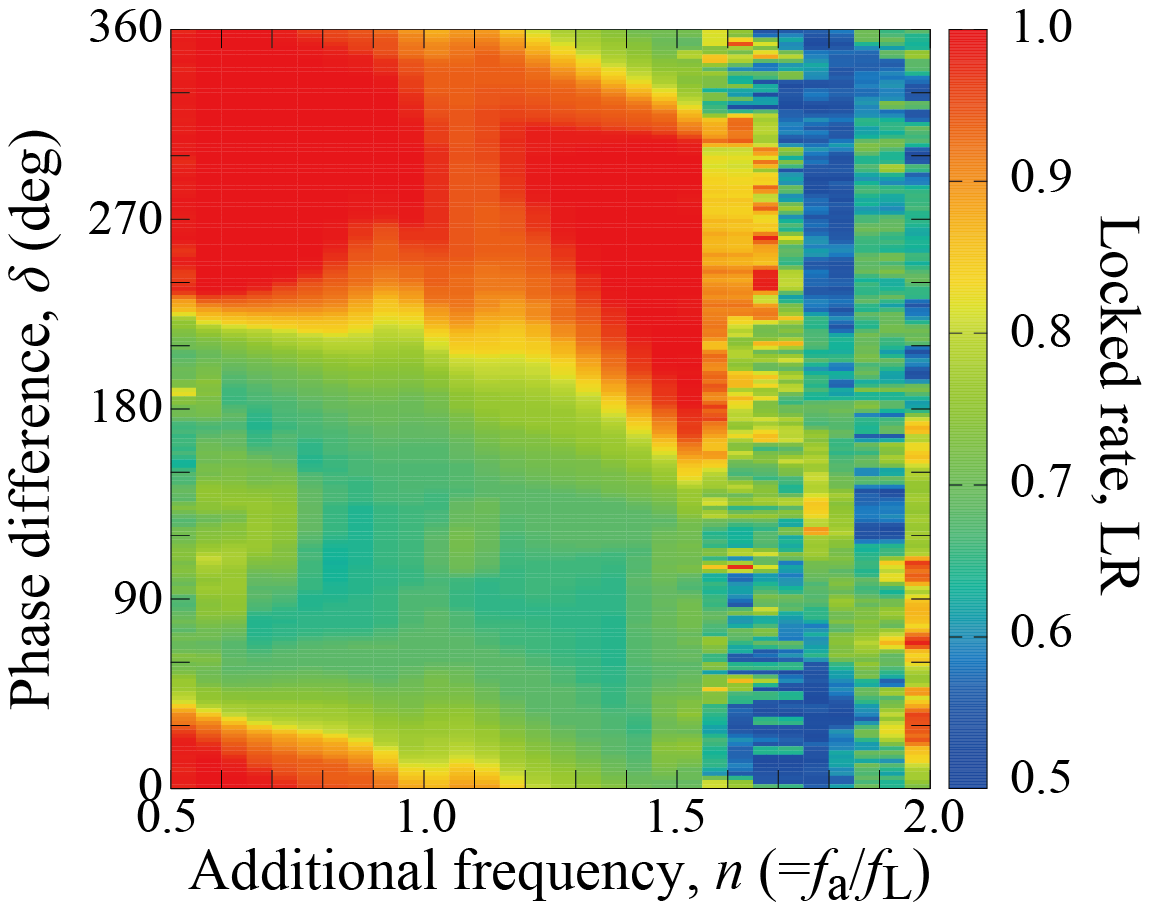}}
\caption{
             Dependence of locked rate (LR) on a frequency ($f_{\rm a}$ in terms of $n=f_{\rm a}/ f_{\rm L}$) and phase ($\delta$) of an additional voltage. 
         \vspace{-3ex}}
\label{fig:fig4}
\end{figure}


We perform similar calculations for various parameters (see also \ref{sec:AppendixC} showing the locked rate for different initial conditions and parameters). 
Figure \ref{fig:fig4} summarizes the dependence of the locked rate on the frequency $f_{\rm a}$ and phase $\delta$ of the additional VCMA effect, where we express $f_{\rm a}$ in terms of a ratio $n=f_{\rm a}/f_{\rm L}$ between $f_{\rm a}$ and the Larmor frequency $f_{\rm L}$. 
We observe a trigonometric-function-like dependence of the locked rate with respect to $\delta$ for $n\lesssim 1.5$. 
The locked rate can be $1$ when $f_{\rm a}$ (or $n=f_{\rm a}/f_{\rm L}$) and $\delta$ are appropriately chosen. 
The locked rate becomes low and random as $n$ becomes close to $2$. 
This is reasonable because the case of $n=2$ corresponds to the original parametric oscillation, i.e., the additional voltage just enhances the amplitude of the original voltage with the frequency $f=2f_{\rm L}$, and thus, the phase is not uniquely locked. 
The results shown in Fig. \ref{fig:fig4} indicates that the additional voltage with a frequency different from $f=2f_{\rm L}$ can manipulate the rate of the phase locking. 
In next section, let us view this result from an analytical study on the LLG equation. 



\section{Analytical treatment}
\label{sec:Analytical treatment}


The LLG equation, Eq. (\ref{eq:LLG}), with Eq. (\ref{eq:H_K}) being adopted is a time-dependent (non-autonomous) nonlinear differential equation. 
Therefore, it is generally difficult to analytically solve \cite{perko01} and provide a quantitative description of the numerical results. 
We will, however, examine to find some insight for the numerical results by applying some approximations to the LLG equation and obtain analytical solution, which will provide a qualitative viewpoint on why the additional VCMA effect contributes to lock the oscillator phase. 
In the following, we first show our investigation of an analytical solution of the LLG equation in a steady state and then provide some comments on the issues remained in the analyses. 


\subsection{Steady state solution of the LLG equation in the presence of additional microwave voltage}
\label{ref:Steady state solution of the LLG equation in the presence of additional microwave voltage}

Since the magnetization in the parametric oscillation state oscillates around the $x$ axis, it is convenient to introduce zenith and azimuth angles, $\Theta$ and $\Phi$ as $\mathbf{m}=(m_{x},m_{y},m_{z})=(\cos\Theta,\sin\Theta\cos\Phi,\sin\Theta\sin\Phi)$. 
For simplicity, we introduce notations $h=[\gamma/(1+\alpha^{2})] H_{\rm appl}$, $k=[\gamma/(1+\alpha^{2})] H_{\rm K}^{(1)}$, and $q=[\gamma/(1+\alpha^{2})] H_{\rm K}^{(2)}$. 
The LLG equation for $\Theta$ and $\Phi$ are given as $d\Theta/dt=-(1/\sin\Theta)(\partial\varepsilon/\partial\Phi)-\alpha(\partial\varepsilon/\partial\Theta)$ and $\sin\Theta(d\Phi/dt)=(\partial\varepsilon/\partial\Theta)-\alpha(1/\sin\Theta)(\partial\varepsilon/\partial\Phi)$, where 
\begin{equation}
  \varepsilon
  =
  -h \cos\Theta
  -
  \frac{k}{2}
  \sin(2\pi ft)
  \sin^{2}\Theta
  \sin^{2}\Phi
  -
  \frac{q}{2}
  \sin(2\pi f_{\rm a}t+\delta)
  \sin^{2}\Theta
  \sin^{2}\Phi. 
\end{equation}
The LLG equation is explicitly written as 
\begin{equation}
\begin{split}
  \frac{d\Theta}{dt}
  =
  &
  \left[
    k \sin(2\pi ft)
    +
    q \sin(2\pi f_{\rm a}t+\delta)
  \right]
  \sin\Theta
  \sin\Phi
  \cos\Phi
\\
  &-
  \alpha
  \left\{
    h
    -
    \left[
      k
      \sin(2\pi ft)
      +
      q 
      \sin(2\pi f_{\rm a}t+\delta)
    \right]
    \cos\Theta
    \sin^{2}\Phi
  \right\}
  \sin\Theta, 
  \label{eq:LLG_theta}
\end{split}
\end{equation}
\begin{equation}
\begin{split}
  \frac{d\Phi}{dt}
  =
  &
  h
  -
  \left[
    k 
    \sin(2\pi ft)
    +
    q
    \sin(2\pi f_{\rm a}t+\delta)
  \right]
  \cos\Theta
  \sin^{2}\Phi
\\
  &+
  \alpha
  \left[
    k 
    \sin(2\pi ft)
    +
    q
    \sin(2\pi f_{\rm a}t+\delta)
  \right]
  \sin\Phi
  \cos\Phi. 
  \label{eq:LLG_phi}
\end{split}
\end{equation}
For the further discussion, it is convenient to introduce a new variable $\Psi=\Phi-\pi ft$, instead of $\Phi$, which is the difference between the phase $\Phi$ of the magnetization and the half of the phase of the microwave voltage. 
In the parametric oscillation state, the phase $\Phi$ changes as $\Phi\propto 2\pi f_{\rm L}t$ and the frequency of the microwave voltage $f$ satisfies $f=2f_{\rm L}$; therefore, $\Psi$ approximately becomes a constant value. 
The numerical simulation also implies that $\Theta(=\cos^{-1}m_{x})$ is approximately constant during the oscillation; see Figs. \ref{fig:fig2}(b) and \ref{fig:fig3}(a). 
Therefore, we average Eqs. (\ref{eq:LLG_theta}) and (\ref{eq:LLG_phi}) over a period $1/f_{\rm L}$ by assuming that $\Theta$ and $\Psi$ are constants and obtain, 
\begin{equation}
\begin{split}
  \overline{
    \frac{d\Theta}{dt}
  }
  =&
  \frac{k}{4}
  \sin\Theta
  \cos 2\Psi
  -
  \alpha
  \left(
    h
    -
    \frac{k}{4}
    \cos\Theta
    \sin2\Psi
  \right)
  \sin\Theta
\\
  &-
  \frac{q [2\cos(n\pi+\delta) \cos2\Psi + n \sin(n\pi + \delta)\sin 2\Psi] \sin n\pi \sin\Theta}{2(4-n^{2})\pi}
\\
  &-
  \frac{\alpha q \{ [-4+n^{2}(1-\cos2\Psi)]\sin(n\pi+\delta) + 2n \cos(n\pi+\delta) \sin 2\Psi \} \sin n\pi \sin\Theta\cos\Theta}{2(4-n^{2})n\pi}, 
  \label{eq:LLG_theta_ave}
\end{split}
\end{equation}
\begin{equation}
\begin{split}
  \overline{
    \frac{d\Psi}{dt}
  }
  =&
  \sigma
  -
  \frac{k}{4}
  \cos\Theta
  \sin 2\Psi 
  +
  \frac{\alpha k}{4}
  \cos 2\Psi
\\
  &+
  \frac{q \{ [-4+n^{2}(1-\cos2\Psi)]\sin(n\pi+\delta) + 2n \cos(n\pi+\delta) \sin 2\Psi \} \sin n\pi \cos\Theta}{2(4-n^{2})n\pi}
\\
  &-
  \frac{\alpha q [2\cos(n\pi+\delta) \cos2\Psi + n \sin(n\pi + \delta)\sin 2\Psi] \sin n\pi}{2(4-n^{2})\pi}, 
  \label{eq:LLG_psi_ave}
\end{split}
\end{equation}
where $\sigma=h-\pi f$, and we introduce $n=f_{\rm a}/f_{\rm L}$, as done in Fig. \ref{fig:fig4}. 
Although $\sigma=0$ for the present work ($f=2f_{\rm L}$), we keep this term ($\sigma$) for the following discussion. 
While the steady-state ($\overline{d\Theta/dt}=0$ and $\overline{d\Psi/dt}=0$) solutions of $\Theta$ and $\Psi$ for $q [\propto H_{\rm K}^{(2)}]=0$ can be analytically obtained \cite{taniguchi24}, it is difficult to extend the solutions for the case of nonzero $q$. 
However, Eq. (\ref{eq:LLG_psi_ave}) still provides a qualitative insight of the role of the additional VCMA effect ($\propto q$) on the phase locking. 
For this purpose, first, let us first consider Eq. (\ref{eq:LLG_psi_ave}) in the limit of $q\to 0$; 
\begin{equation}
  \overline{
    \frac{d\Psi}{dt}
  }
  =
  \sigma
  -
  \frac{k}{4}
  \cos\Theta
  \sin 2\Psi
  +
  \frac{\alpha k}{4}
  \cos 2\Psi. 
\end{equation}
This equation implies that the variable $\Psi$ moves in a potential 
\begin{equation}
  U
  =
  -\sigma 
  \Psi
  -
  \frac{k}{8}
  \cos\Theta
  \cos 2\Psi 
  -
  \frac{\alpha k}{8}
  \sin 2\Psi, 
  \label{eq:potential}
\end{equation}
as $\overline{d\Psi/dt}=-\partial U/\partial \Psi$. 
When the frequency $f$ of the microwave VCMA effect is twice the Larmor frequency $f_{\rm L}$ and thus, the condition of the parametric oscillation ($\sigma=0$) is satisfied, the potential $U$ is described by trigonometric functions, $\cos2\Psi$ and $\sin 2\Psi$, of $2\Psi$. 
Thus, the potential $U$ have two minima, and their depths are the same. 
It means that the phase in the parametric oscillation states has two possible values, and realizing which of the states depends on the initial state. 
On the other hand, for a finite $\sigma$, i.e., when the condition of the parametric oscillation is slightly broken, the potential becomes asymmetric due to the term $-\sigma \Psi$ in Eq. (\ref{eq:potential}), and one of two possible states is favored than the other \cite{taniguchi23}. 


The role of the additional VCMA effect [$\propto q(\neq 0$)] might be understood in a similar way. 
Equation (\ref{eq:LLG_psi_ave}) indicates the presence of a term proportional to $q (4-n^{2}) \cos\Theta\sin n\pi \sin(n\pi +\delta)$, which is independent of $\Psi$. 
Therefore, this term plays a similar role to $\sigma$ and provides an asymmetry in the potential, which leads to the phase locking of the parametric oscillator. 
The fact that this term includes a trigonometric function of $\delta$ also agrees with the numerical results in Fig. \ref{fig:fig3}(c) that the locked rate depends on the phase $\delta$ and the dependence is trigonometric-function-like. 
Therefore, the above analysis provides a viewpoint for clarifying the role of the additional VCMA effect on the phase locking. 


\begin{figure}
\centerline{\includegraphics[width=1.0\columnwidth]{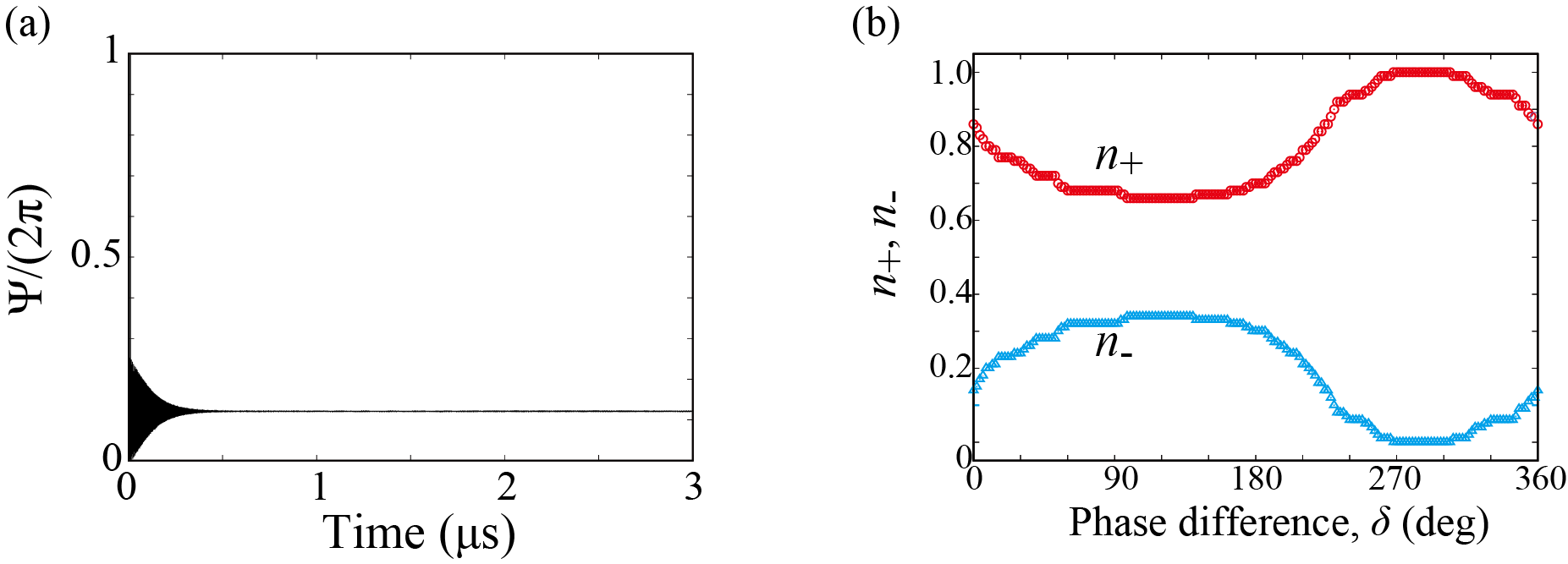}}
\caption{
             (a) An example of the numerically evaluated time evolution of $\Psi$. 
             (b) Dependence of $n_{\pm}$ as a function of $\delta$. 
             The values of the parameters are identical to those used in Fig. \ref{fig:fig3}. 
         \vspace{-3ex}}
\label{fig:fig5}
\end{figure}


At the end of  this subsection, we investigate the asymmetry of the potential by the additional VCMA effect by the following way. 
Recall that we evaluated the locked rate, defined by Eq. (\ref{eq:locked_rate}), numerically in Sec. \ref{sec:Numerical simulation}. 
There, we introduced $N_{+}$ and $N_{-}$, which are numbers of the samples having one of two possible phases. 
Note that there was not solid distinction between $N_{+}$ and $N_{-}$; for example, we did not define which of two solid lines in Fig. \ref{fig:fig2}(d) belongs to $N_{+}$ (or $N_{-}$). 
This was because there was no standard on the origin of the phase. 
This is a common issue for any oscillating systems; when there are two oscillators with the same frequency, their phases are not uniquely determined unless the origin is not specified. 
This is also true for the present system. 
However, the analytical treatment developed above implies a possibility to define $N_{+}$ and $N_{-}$ uniquely. 
The fact that the potential $U$ consists of trigonometric functions, $\cos2\Psi$ and $\sin 2\Psi$, means that the potential has two minima, and one of them locates in the region of $0 \le \Psi < \pi$ while the other locates in the region of $\pi \le \Psi < 2\pi$. 
Therefore, let us define $N_{+}$ and $N_{-}$ as the number of the samples whose $\Psi$ in a steady state locates in $0 \le \Psi < \pi$ and $\pi \le \Psi < 2\pi$, respectively. 
For example, Fig. \ref{fig:fig5}(a) shows an example of the time evolution of $\Psi$, where the values of the parameters are identical to those used in Fig. \ref{fig:fig3}. 
In this example, $\Psi$ saturates to the region of $0 \le \Psi < \pi$, and thus, this example belongs to $N_{+}$. 
After defining $N_{+}$ and $N_{-}$ as such, we introduce 
\begin{equation}
  n_{+(-)}
  =
  \frac{N_{+(-)}}{N_{+}+N_{-}}. 
  \label{eq:n_def}
\end{equation}
Note that $\Psi$ is not a quantity directly measured in the experiments; rather, for example, $\Phi$ is measured directly because it determines the oscillation of the output power. 
We, however, mention that $\Psi$ will be experimentally evaluated by applying a post-processing to the measured data. 
Therefore, it will be meaningful to the experimental researchers to introduce $n_{\pm}$ in Eq. (\ref{eq:n_def}). 
Figure \ref{fig:fig5}(b) shows the dependence of $n_{\pm}$ on $\delta$. 
Here, we notice that $n_{+}$ is identical to the locked rate shown in Fig. \ref{fig:fig3}(c). 
This fact supports the analyses developed above, i.e., the additional VCMA effect makes the potential $U$ asymmetric and makes one of two possible phases more realizable; in the case of Fig. \ref{fig:fig5}(b), $n_{+}$ becomes more realizable, and thus, is identical to the locked rate.



\subsection{Remained issues}
\label{sec:Remained issues}

We should admit that the analytical description developed above does not predict the phase locking when $n$ is an integer (or equivalently, when the frequency $f_{\rm a}$ of the additional voltage is an integer multiple of the Larmor frequency), in contrast to the result shown in Fig. \ref{fig:fig3}(c). 
The contradiction may possibly come from the assumption used during the averaging of Eqs. (\ref{eq:LLG_theta_ave}) and (\ref{eq:LLG_psi_ave}). 
While we assume that $\Theta$ is approximately constant, the numerical simulation indicates that $m_{x}=\cos\Theta$ slightly oscillates, as shown in Figs. \ref{fig:fig2}(b) and \ref{fig:fig3}(a). 
This approximation was necessary to develop the analytical treatment of the LLG equation but may possibly lead to differences between the numerical and analytical results. 
It is, however, difficult to solve the LLG equation analytically without using this assumption, even in the case of $q= 0$ \cite{taniguchi24}. 
We consider overcoming the approximation as future work. 


We also admit that the theoretical analyses developed here do not clarify the dependence of the solution of the LLG equation on the initial condition. 
The above analyses investigate the steady state solutions only. 
While the steady state solutions indicate the presence of two possible values of $\Psi$, which of them is realized cannot be found from the solutions, i.e., it is unclear how the initial condition relates to the value of $\Psi$; see also \ref{sec:AppendixC}, where locked rates for different initial conditions are shown. 
It is currently difficult to reveal the dependence of the solution of the LLG equation on the initial condition analytically, even in a steady state, because the LLG equation is a time-dependent (non-autonomous) nonlinear differential equation. 
We would also like to keep this issue as a future work. 



\section{Conclusion}
\label{sec:Conclusion}

In conclusion, the phase locking of a voltage-controlled parametric oscillator by applying an additional microwave VCMA effect was investigated by numerically solving the LLG equation. 
With wide range of the frequency and phase of the additional voltage, the phase locking was confirmed from the simulation. 
The locked rate of the oscillator phase shows a trigonometric-function-like dependence on the phase of the additional voltage. 
The result was qualitatively explained by an analytical treatment of the LLG equation. 
The method will provide a tool to manipulate the phase of the voltage-controlled parametric oscillator, where the phase locking is required for applications such as brain-inspired computing while the phase unlocking is desirable for other applications such as random-number generators. 


\section*{Data availability}

Data will be made available on request. 

\section*{Acknowledgement} 

The work is supported by JSPS KAKENHI Grant Number 20H05655 and 24K01336. 
The author is grateful to Takayuki Nozaki and Takehiko Yorozu for valuable discussion. 


\appendix


\section{Sign of perpendicular magnetic anisotropy field}
\label{sec:AppendixA}

In this Appendix, we provide some comments on materials and architectures on the VCMA devices, although it was partially written in Sec. \ref{sec:System description}. 
The VCMA effect in metallic systems have been investigated experimentally by using magnetic tunnel junctions (MTJs) consisting of ferromagnetic free layers, which are typically Co-Fe-B \cite{yamamoto20} or Fe \cite{,maruyama09,shiota09,nozaki12,nozaki19}, and MgO insulating layer [4,30-33]. 
The shape of the films in recent experiments is a cylinder for realizing high-density non-volatile memory. 
The parameters used in our simulations were brought from these experiments, where their specific values were already written in Sec. \ref{sec:System description}. 
In the following, we provide additional comments on the material parameters and architectures on the VCMA devices. 

The magnitude of the shape magnetic anisotropy field along the $z$ direction is given by $4\pi M$ and is about $1$-$2$ T for typical Co-Fe-based magnetic tunnel junctions, where the saturation magnetization is about $1000$-$1500$ emu/cm${}^{3}$ and a demagnetization coefficient in the $z$ direction is close to $1$ due to thin-flim structure. 
The reduction of net demagnetization field by the interfacial perpendicular magnetic anisotropy field in magnetic tunnel junctions was reported in Ref. \cite{yakata09}, where the sign of the perpendicular magnetic anisotropy field was still negative ($H_{\rm K}<0$) and thus, the in-plane magnetized state was stable. 
The perpendicularly magnetized state by the interfacial effect was reported in Ref. \cite{ikeda10} after one year from Ref. \cite{yakata09}, but a similar result was already found in Ref. \cite{shinjo79} many decades ago. 

The VCMA effect provides an additional modulation to the perpendicular magnetic anisotropy field. 
In contrast to the interfacial contribution, which is determined by the fabrication process of the magnetic tunnel junctions and cannot be changed after the process, the VCMA effect enables us to modulate the perpendicular magnetic anisotropy field even after the sample is fabricated. 
The magnitude of the perpendicular magnetic anisotropy field by the VCMA effect is given as $(2\eta V_{\rm appl})/(Mdd_{\rm I})$, where $\eta$, $V_{\rm appl}$, and $d_{\rm I}$ are the VCMA efficiency, applied voltage, and thickness of the insulating layer. 
Note that this equation assumes an accumulation of electrons due to the insulating layer; therefore, although it looks that the VCMA effect increases as the thickness $d_{\rm I}$ of the insulating layer becomes thin, the thickness of the insulating layer used for VCMA devices should be thick enough to increase resistance of the device and to  prevent electrons' flow. 
In fact, for example, the resistance of a VCMA device in Ref. \cite{shiota17} is at least one or two orders of magnitude larger than that of spin-torque oscillator in Ref. \cite{kubota13}. 
The limit of $d_{\rm I}\to 0$ means that the system becomes metallic, and the VCMA effect will disappear. 
Therefore, the value of $d_{\rm I}$ cannot be excessively thin and should be at least about $1.5$-$2.0$ nm thick for typical experiments. 
On the other hand, the thickness $d$ of the free layer should be designed thin for enhancing the VCMA effect. 
The VCMA efficiency $\eta$ found in experiments is about 350 fJ/(V m) at maximum \cite{nozaki20}. 
The magnitude of the voltage $V_{\rm appl}$ available in experiments depends on its pulse width, as well as the resistance of the device, to avoid electrostatic breakdown of the magnetic tunnel junctions.
It can be $3.0$ V for an ultra-short pulse, while it is about $1.0$ at maximum in the experiment of the parametric oscillation \cite{yamamoto20}. 
Accordingly, the perpendicular magnetic anisotropy field caused by the VCMA effect is on the order of kilo Oersted at maximum. 
Therefore, the values of $H_{\rm K}^{(1)}$ and $H_{\rm K}^{(2)}$ used in the present numerical simulation (on the order of $100$ Oe) are reasonable. 
In addition, regarding the interfacial contribution mentioned above, the sign of net perpendicular magnetic anisotropy field can either be negative or positive by  the VCMA effect. 
This fact is used not only in the non-volatile memory applications \cite{nozaki19} but also in other applications, such as the parametric oscillator \cite{yamamoto20} and associative memory operation \cite{taniguchi24SR}. 



\section{Thermal fluctuation in numerical simulation}
\label{sec:AppendixB}

In this work, the LLG equation is solved by the $4^{\rm th}$ order Runge-Kutta method with a time increment $\Delta t =1$ ps. 
The random field $h_{a}(t)$ used in the preparation of the initial state is defined in the numerical simulation as 
\begin{equation}
  h_{a}
  =
  \sqrt{
    \frac{2\alpha k_{\rm B}T}{\gamma MV \Delta t}
  }
  \xi_{a}, 
\end{equation}
where a random number $\xi_{a}$ is defined from uniform random numbers $\zeta_{a}$, $\zeta_{b}$ ($0<\zeta_{a}, \zeta_{b}<1$) by the Box-Muller transformation, $\xi_{a}=\sqrt{-2\ln \zeta_{a}}\cos(2\pi \zeta_{b})$ and $\xi_{b}=\sqrt{-2\ln \zeta_{a}}\sin(2\pi \zeta_{b})$. 
Remind that the random field is added to the LLG equation only when the initial state is prepared to reveal the role of the additional voltage in the phase locking clearly. 


\section{Numerical results for different conditions}
\label{sec:AppendixC}

In this Appendix, we show the locked rates for different initial conditions and parameters.


\begin{figure}
\centerline{\includegraphics[width=1.0\columnwidth]{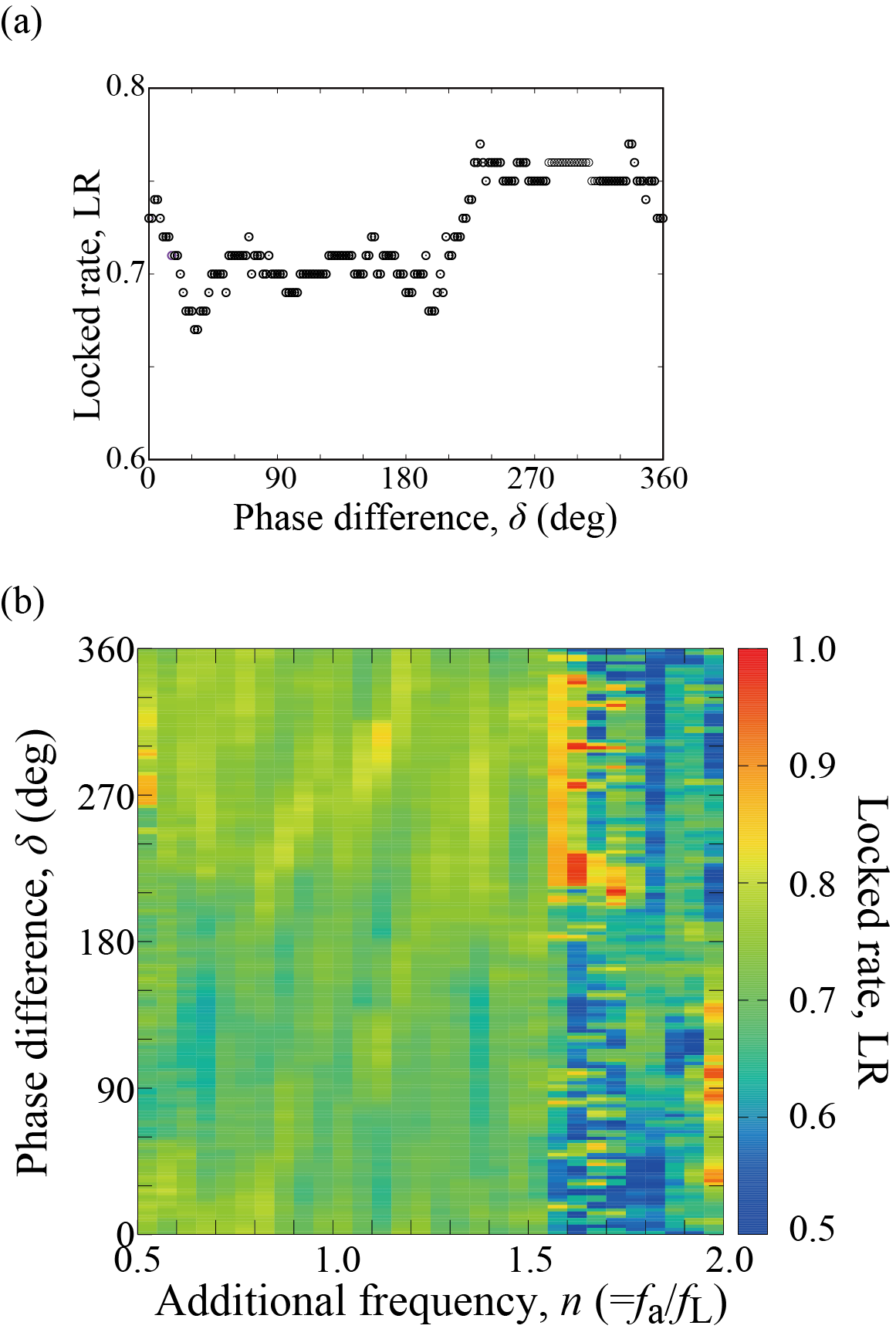}}
\caption{
            (a) Locked rate as a function of $\delta$ for $H_{\rm K}^{(2)}=50$ Oe and $f_{\rm a}=f_{\rm L}$, where the initial state is close to the negative $z$ direction. 
            (b) Dependence of the locked rate on $n(=f_{\rm a}/f_{\rm L})$ and $\delta$ when the initial sate is close to the negative $z$ direction. 
         \vspace{-3ex}}
\label{fig:fig6}
\end{figure}


First, we show the locked rate when the initial condition is close to the negative $z$ direction; recall that the initial state in the main text were close to the positive $z$ direction. 
In this case, we first set $\mathbf{m}$ to the negative $z$ direction and prepared $100$ initial conditions by adding the thermal fluctuation, as done in Sec. \ref{sec:System description}. 
Figure \ref{fig:fig6}(a) shows the dependence of the locked rate on the phase of the additional voltage $\delta$, where the values of the parameters are identical to those used in Fig. \ref{fig:fig3}(c). 
We still observe a trigonometric-function-like dependence of the locked rate on $\delta$, however, the dependence is relatively weak compared to the result shown in Fig. \ref{fig:fig3}(c). 
We also investigate the locked rate for the different frequencies $f_{\rm a}$ of the additional voltage and again find relatively weak trigonometric-function-like dependence; see Fig. \ref{fig:fig6}(b). 
We also evaluate the locked rate by setting the initial state close to the positive $x$ direction (parallel to the applied magnetic field, $H_{\rm appl}$), and find that the locked rate is almost $0.5$ for any $\delta$. 
We do not think that the weak dependence of the locked rate on $\delta$ when the initial condition of $\mathbf{m}$ is close to the negative $z$ direction reflects any asymmetry between the positive and negative initial states because the LLG equation does not include any asymmetry with respect to the $z$ direction. 
Instead, these results imply that the locked rate strongly depends on the initial conditions of the magnetization; 
if we use different $100$ initial conditions, the results might be changed. 
As mentioned in Sec. \ref{sec:Introduction}, the motivation of the present work is to find a way to manipulate the locked rate by overcoming this issue. 
Our numerical results indicate that the additional VCMA effect partially solves the issue, however, other solutions will be still necessary. 


\begin{figure}
\centerline{\includegraphics[width=1.0\columnwidth]{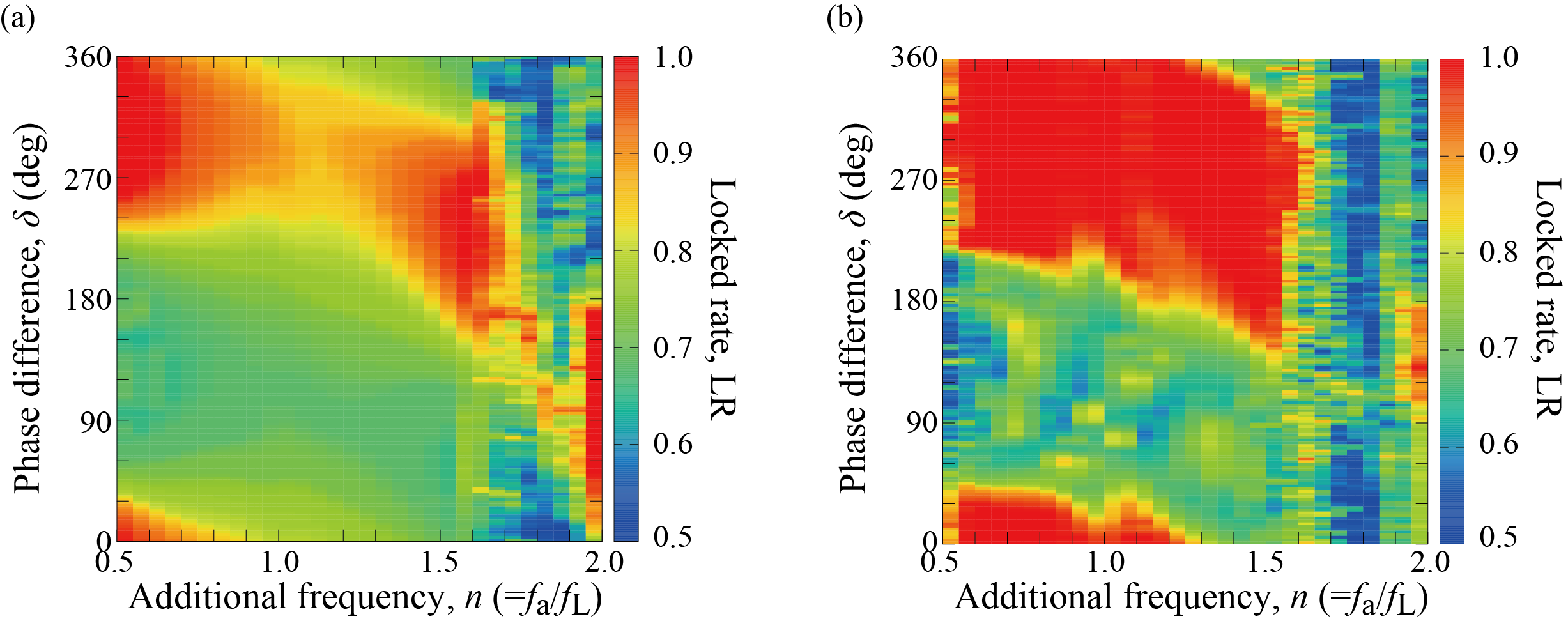}}
\caption{
            Dependence of locked rates on $n(=f_{\rm a}/f_{\rm L})$ and $\delta$ of an additional voltage, where $H_{\rm K}^{(2)}$ is (a) $25$ Oe and (b) 100 Oe, while $H_{\rm K}^{(1)}$ is $300$ Oe. 
         \vspace{-3ex}}
\label{fig:fig7}
\end{figure}


We also evaluate the locked rate for the different values of $H_{\rm K}^{(2)}$. 
Figures \ref{fig:fig7}(a) and \ref{fig:fig7}(b) show the locked rate for the $H_{\rm K}^{(2)}$ of (a) $25$ Oe and (b) $100$ Oe; recall that it was $50$ Oe in Fig. \ref{fig:fig4}. 
The other parameters, as well as the initial conditions, are identical to those used in Fig. \ref{fig:fig4}.  
We consider that changing the value of $H_{\rm K}^{(2)}$ in this range is reasonable to clarify the role of the additional VCMA effect on the locked rate.
This is because, while $H_{\rm K}^{(2)}$ can be on the order of kilo Oersted at maximum, the realization of such a large $H_{\rm K}^{(2)}$ is possible only for a short pulse, as mentioned in \ref{sec:AppendixA}. 
Thus, the experimentally available value for the parametric oscillation will be on the order of $10$-$100$ Oe for current technology, although a rapid growth of the VCMA efficiency \cite{nozaki20} will make it possible soon to realize larger $H_{\rm K}^{(2)}$. 
Comparing Fig. \ref{fig:fig7}(a) with Fig. \ref{fig:fig4}, we observe that the locked rate increases as $H_{\rm K}^{(2)}$ increases. 
On the other hand, comparing Fig. \ref{fig:fig7}(b) with Fig. \ref{fig:fig4}, the locked rate is almost saturated. 
Therefore, we consider that the result in Fig. \ref{fig:fig4} appropriately shows the ability of the additional VCMA effect on the phase locking. 








\end{document}